\documentclass[aps,pre,showpacs,amsmath,amssymb,superscriptaddress,twocolumn]{revtex4-1}

\usepackage{graphicx}
\usepackage{amsmath}
\usepackage{hyperref}
\usepackage{dcolumn}
\usepackage{bm}
\usepackage{color}
\usepackage{nicefrac}

\begin{document}

\title{Condensation of Lee-Yang zeros in scalar field theory }

\author{N.~G.~Antoniou}
 \email[]{nantonio@phys.uoa.gr}
\affiliation{Faculty of Physics, University of Athens, GR-15784 Athens, Greece}

\author{F.~K.~Diakonos}
 \email[]{fdiakono@phys.uoa.gr}
\affiliation{Faculty of Physics, University of Athens, GR-15784 Athens, Greece}

\author{X.~N.~Maintas}
 \email[]{xmaintas@phys.uoa.gr}
\affiliation{Faculty of Physics, University of Athens, GR-15784 Athens, Greece}

\author{C.~E.~Tsagkarakis}
 \email[]{ctsagkarakis@phys.uoa.gr}
\affiliation{Faculty of Physics, University of Athens, GR-15784 Athens, Greece}

\date{\today}

\begin{abstract}
\noindent
We show that, at the critical temperature, there is a class of Lee-Yang zeros of the partition function in a general scalar field theory, which location scales with the size of the system with a characteristic exponent expressed in terms of the isothermal critical exponent $\delta$. In the thermodynamic limit the zeros belonging to this class condense to the critical point $\zeta=1$ on the real axis in the complex fugacity plane while the complementary set of zeros (with $\Re {\zeta} < 1$) covers uniformly the unit circle. Although the aforementioned class degenerates to a single point for an infinite system, when the size is finite it dominates in the partition function and determines the self-similar structure (fractal geometry, scaling laws) of the critical system. This property opens up the perspective to formulate finite-size scaling theory in effective QCD, near the chiral critical point, in terms of the location of Lee-Yang zeros. 
\end{abstract}


\maketitle

The zeros of the partition function in the complex control parameter plane should provide a valuable tool for gaining insight into critical phenomena, as originally proposed by Lee and Yang in their seminal work about 65 years ago \cite{Lee1952}. Since then, the Lee-Yang scenario has been confirmed and/or generalized studying the distribution of the partition function complex zeros in a variety of models ranging from classical Ising spins on the lattice to non-equilibrium classical and quantum phase transitions \cite{Byckling1965,Abe1967,Asano1968,Grossmann1968,Suzuki1970,Ruelle1971,Kortman1971,Newman1974,Fisher1978,Lieb1981,Yamada1981,Itzykson1983,Ozeki1988,Kenna1994,
Matveev1996,Creswick1997,Simon1997,Biskup2000,Blythe2003,Janke2004,Bena2005,Kim2006,Janke2006,Glumac2013,Hickey2013,Hickey2014}. Furthermore, experimental verifications of Lee-Yang theory (LYT) using isothermal magnetization data of FeCl$_2$ \cite{Binek1998,Binek2001} or measuring quantum coherence of a probe spin coupled to an Ising spin-bath  \cite{Peng2015} have also been reported in the last two decades.

In some cases such as monomer-dimer systems \cite{Heilmann1972} or complex networks of Ising spins \cite{Lebowitz2012,Krasnytska2015} the LYT is found to be violated through the occurrence of an accumulation of zeros on the negative real axis \cite{Lebowitz2016} instead of the unit circle.  Recently, the LYT has been implemented to analyze the phase diagram of Quantum Chromodynamics (QCD) on the lattice \cite{Barbour1992,Fodor2002,Fodor2004,Ejiri2006}. The idea is that, extending analytically the partition function on the complex plane of an involved control parameter, the associated zeros move, with increasing system size, differently within this plane in the case of a crossover compared to a first or second order transition. However the technical complexity of the theory does not allow to obtain a clear picture and effective models such as random matrix models \cite{Stephanov2006,Morita2015} or the reconstruction of the partition function directly from experimental data \cite{Nakamura2016} have been employed to shed light on this challenging problem.

Despite these extensive theoretical and experimental investigations the ultimate goal of LYT, i.e. the deeper understanding of critical phenomena in terms of the zeros of the partition function, is still pending. In the present Letter we perform a step in this direction exploring properties of the set of the partition function zeros in the complex fugacity ($\zeta$) plane for a scalar field at the critical temperature. We show that the set of these zeros is divided in two subsets with different scaling properties. One subset exists only for finite system's size and  degenerates to the point $\zeta_c=(1,0)$ in the thermodynamic limit while the complementary set of complex zeros survives in the thermodynamic limit covering uniformly the unit circle in the complex fugacity plane. The first set is shown to curry the information for the self-similar structure of the critical system, while the second set is not directly related to criticality. The distance from $\zeta_c$ of the partition function complex zeros, belonging to the first set, scales with the size of the system with an exponent $q$ depending on the isothermal critical exponent $\delta$. This exponent is in fact $q=d_F/d$ where $d_F$ is the fractal mass dimension characterizing the self-similar critical system and $d$ is the embedding dimension. 
Our analysis clearly demonstrates the power of LYT when applied to finite systems and opens up the perspective to employ the location of partition function zeros for a rigorous finite size scaling analysis of the critical properties of effective QCD close to the chiral transition point.  

We start our analysis writing the general form for the partition function of a thermodynamic system in the grand-canonical ensemble:
\begin{equation}
\mathcal{Z}(\zeta,V,T)={\displaystyle{\sum_{N=0}^M}} \zeta^N~ {\displaystyle{\exp[-\beta F(N,V,T)]}}
\label{eq:1}
\end{equation}
where $\zeta=\exp(\frac{\mu-\mu_c}{k_B T})$ ($\mu$ = chemical potential, $T$ = temperature), $F$ is the Helmholtz free energy and $M$ the maximum number of particles in volume $V$. We have assumed a hard-core interaction in the fluid, introducing a length scale $V_0^{1/d}$ ($M=V/V_0$). In the following, without loss of generality, we will set $d=3$ for the embedding dimension. A critical system described by Eq.~(\ref{eq:1}) becomes self-similar at the critical point $(\zeta_c=1,~T=T_c)$ with fractal dimension $d_F=\frac{3 \delta}{\delta +1}$ linked to the isotherm critical exponent $\delta$ which is associated with the order parameter $N$. In the thermodynamic limit ($V \gg V_0$) the related critical fluctuations are described by the geometrical law:
\begin{equation}
\langle N \rangle \sim  M^q~~~~~~;~~~~~~ q=d_F/3,~M=\frac{V}{V_0} \to \infty
\label{eq:2}
\end{equation}
with $q < 1$. This distinct property of the critical system implies a scaling hypothesis for the free energy $F(N,M,T_c)$ in Eq.~(\ref{eq:1}):
\begin{equation}
F(N,M,T_c)=\tilde{F}_c(\frac{N}{M^q})~~~~~;~~~~~q < 1,~~M \to \infty
\label{eq:3}
\end{equation}
In fact it is straightforward to show that Eq.~(\ref{eq:3}) is a necessary and sufficient condition for the development of the fractal structure described by Eq.~(\ref{eq:2}) in the thermodynamic limit. In the following we provide a sketch of the proof.

\noindent
{\em (i) Sufficient condition}. The grand-canonical partition function $\mathcal{Z}_M$ at the critical point $(T=T_c,\zeta=\zeta_c=1)$ is given by:
\begin{equation}
\mathcal{Z}_{M,c}=\mathcal{Z}(1,M,T_c)={\displaystyle{\sum_{N=0}^M}} \exp[-\beta_c {\displaystyle{\tilde{F}_c({\displaystyle{\frac{N}{M^q}}})}}]
\label{eq:4}
\end{equation}
Introducing the variable $\chi=N M^{-q}$ we can write Eq.~(\ref{eq:4}) in the thermodynamic limit ($M \to \infty$) as the integral:
\begin{equation} 
\mathcal{Z}_{M,c}=M^q {\displaystyle{\int_0^{M^{1-q}}}} d\chi~ \exp[-\beta_c {\displaystyle{\tilde{F}_c(\chi)}}]
\label{eq:5}
\end{equation}
which leads to 
\begin{equation} 
\mathcal{Z}_M \sim M^q 
\label{eq:6}
\end{equation}
for $M \to \infty$ if $q < 1$. Then, from the defining relation:
\begin{equation}
\langle N \rangle =\zeta \frac{\partial}{\partial \zeta} \ln \mathcal{Z}_M \mid_{\zeta=1}
\label{eq:7}
\end{equation}
we obviously obtain Eq.~(\ref{eq:2}).\\

\noindent
{\em (ii) Necessary condition}. We assume that Eq.~(\ref{eq:2}) holds and we derive Eq.~(\ref{eq:1}). To achieve this, we write the partition function at the critical point as:
\begin{equation}
\mathcal{Z}_{M,c}={\displaystyle{\sum_{N=0}^M}} {\displaystyle{\exp[-\beta_c F(N,M,T_c)]}}
\label{eq:8}
\end{equation}
and for $M \gg 1$ becomes the integral:
\begin{equation}
\mathcal{Z}_{M,c}=M^q {\displaystyle{\int_0^{M^{1-q}}}} d\chi ~ \exp[-\beta_c {\displaystyle{F(\chi M^q,M,T_c)}}]
\label{eq:9}
\end{equation}
where we have used the previously defined variable $\chi$. Since the scaling law $\langle N \rangle \sim M^q$ directly implies that $\mathcal{Z}_{M,c} \sim M^q$, the integral in Eq.~(\ref{eq:9}) must be independent of $M$ for $M \to \infty$. This in turn means that the free energy $F(\chi M^q,M,T_c)$ depends only on $\chi$:
$F(\chi M^q,M,T_c)=\tilde{F}_c(\chi)$ which proves the validity of Eq.~(\ref{eq:1}). 

Thus, the most general form of the partition function at $T=T_c$ which incorporates the critical fluctuations at $\zeta=1$, is written:
\begin{equation}
\mathcal{Z}_M={\displaystyle{\sum_{N=0}^M}} \zeta^N~ {\displaystyle{\exp[-{\displaystyle{f}}({\displaystyle{\frac{N}{M^q};T_c}})]}}
\label{eq:10}
\end{equation}
In what follows we claim that there is a subset of zeros of $\mathcal{Z}_M$ in the complex $\zeta$-plane (Lee-Yang zeros) which in the thermodynamic limit $M \to \infty$ condense at the critical point $\zeta_c=(1,0)$ following a power-law linked to the critical index $q$. To see this, one can write the $M$-th order $\zeta$-polynomial in Eq.~(\ref{eq:10}) in terms of its zeros $\rho_j(M)$ ($j=1,..,M$) and express $\langle N \rangle$ at $\zeta=1$ as:
\begin{equation}
\langle N \rangle = {\displaystyle{\sum_{j=1}^M}} \frac{1}{1-\rho_j(M)} 
\label{eq:11}
\end{equation}
We observe that the critical behaviour $\langle N \rangle \sim M^q$ can be recovered by a simple mechanism of accumulating zeros at $\zeta=1$ for $M \to \infty$:
\begin{equation}
\rho_j(M)=1+\gamma_j M^{-q}
\label{eq:12}
\end{equation}
leading to:
\begin{equation}
\langle N \rangle = M^q {\displaystyle{\sum_{j=1}^M}} \frac{\gamma_{j,R}}{\vert \gamma_j \vert^2}~~~~~;~~~~~M \to \infty
\label{eq:13}
\end{equation}
provided that the series is convergent. In fact only a subset $\{\gamma_l\}$ of the complex numbers $\gamma_j=\gamma_{j,R} + \mathbf{i} \gamma_{j,I}$ ($j=1,M$) is independent of $M$ while the remaining part of the set may depend on $M$ in a manner that is compatible with the convergence of the sum in Eq.~(\ref{eq:13}). Thus, we argue that there is a strong link of the scaling hypothesis ({\em fractality}) at the critical point to the {\em condensation law} (Eq.~(\ref{eq:12})) of a subset of the Lee-Yang zeros in the thermodynamic limit.  

The arguments in the above discussion are further clarified if we consider specific theories. Here we will focus on a class of self-interacting scalar field theories in $3d$ with a stabilizing term $\sigma^{\delta + 1}$ for two reasons: (i) this class is related to the description of a tricritical point ($\delta = 5$) for which the characteristic dimensionality is $d_c=3$ and therefore Landau theory is correct up to logarithmic corrections \cite{Pfeuty1977} and (ii) it is of particular importance for the description of critical fluctuations (at $T=T_c$, $\mu=\mu_c$) in strongly interacting matter, associated with the QCD critical point \cite{Antoniou2006}. At $T=T_c$ the partition function of this system is given as: 
\begin{equation}
\mathcal{Z}(V,T_c)=\displaystyle{\int_{\{\sigma \}}} \mathcal{D}[\sigma]~\exp[-{\displaystyle{\beta_c \int_V d^3\vec{x} H_{c}[\sigma]}}]
\label{eq:14}
\end{equation}
with $\sigma(\vec{x})$ the scalar field, $\beta_c=1/(k_B T_c)$ and
\begin{equation}
H_{c}=\frac{1}{2} (\vec{\nabla} \sigma)^2 + U_{c}[\sigma]~~~;~~~U_{c}[\sigma]=\frac{m^2 \sigma^2}{2}   + \frac{\lambda \sigma^{\delta + 1}}{\delta +1} 
\label{eq:15}
\end{equation}
In Eq.~(\ref{eq:15}) $m^2$ depends on the chemical potential as $m^2 \approx c (\mu - \mu_c) + ~ \textrm{higher~order~terms}$, with $c > 0$, while $\delta$ is the isothermal critical exponent and $\lambda$ an effective coupling with dimension [length]$^{\delta - 3}$. Eq.~(\ref{eq:14}) can be brought to the form of Eq.~(\ref{eq:10}) if we use constant scalar field configurations to saturate the occurring functional integral and switch from field to particle description setting:
\begin{equation}
\int_V \sigma^2~d^3\vec{x} = N V_0^{1/3}~~~~;~~~~N=0,~1,...
\label{eq:16}
\end{equation}
In Eq.~(\ref{eq:16}) $V_0^{1/3}$ is the hard core length scale and the maximum number of scalar particles $M$ is set by the volume $V$ of the system in units of $V_0$: $M=\frac{V}{V_0}$. Within this framework the scalar field $\sigma$ becomes discrete:
\begin{equation}
\sigma^2_N=N \left(\frac{V_0^{1/3}}{V}\right)~~~~~;~~~~~N=0,~1,...,~M
\label{eq:17}
\end{equation}
Defining the fugacity $\zeta$ as:
\begin{equation}
\zeta=\exp[-\frac{1}{2}\beta_c m^2 V_0^{1/3}]
\label{eq:18}
\end{equation}
and inserting Eqs.~(\ref{eq:17},\ref{eq:18}) in Eq.~(\ref{eq:14}) we obtain for the partition function at $T=T_c$ the form:
\begin{equation}
\mathcal{Z}_M=\displaystyle{\sum_{N=0}^M} \zeta^N \exp[-\alpha_c \frac{N^{(\delta + 1)/2}}{M^{(\delta - 1)/2}}]~;~\alpha_c=\frac{\beta_c \lambda}{(\delta +1) V_0} 
\label{eq:19}
\end{equation}
In the thermodynamic limit ($M \to \infty$) the sum in the partition function (\ref{eq:19}) becomes an integral:
\begin{equation}
\mathcal{Z}_M=M \int_0^1~dx~\exp[M(x \ln \zeta - \alpha_c x^{(\delta + 1)/2})]
\label{eq:20}
\end{equation}
Then it is straightforward to calculate the mean number of scalar particles $\langle N \rangle =\zeta \frac{\partial}{\partial \zeta} \ln \mathcal{Z}(\zeta,M)$ as a function of the size of the system $M$ given by:
\begin{equation}
\langle N \rangle = \frac{M  \int_0^1~dx~x~\exp[M(x \ln \zeta - \alpha_c x^{(\delta + 1)/2})]}
{\int_0^1~dx~\exp[M(x \ln \zeta - \alpha_c x^{(\delta + 1)/2})]}
\label{eq:21}
\end{equation}
At the critical point $\zeta_c=(1,0)$ the integrals can be performed analytically leading to:
\begin{equation}
\langle N \rangle = M^{\frac{\delta - 1}{\delta +1}} \left(\frac{1}{\alpha_c}\right)^{2/(\delta +1)} \frac{\gamma(\frac{4}{\delta +1}, \alpha_c M)}{\gamma(\frac{2}{\delta +1}, \alpha_c M)}
\label{eq:22}
\end{equation}
with $\gamma(\nu,x)$ being the incomplete gamma function. For $\alpha_c M \gg 1$ Eq.~(\ref{eq:22}) implies the scaling:
\begin{equation}
\langle N \rangle \sim M^{\frac{\delta - 1}{\delta +1}}
\label{eq:23}
\end{equation}
which in turn suggests the formation of scalar-particle clusters with fractal mass dimension
\begin{equation}
d_F=\frac{3 (\delta -1)}{\delta + 1}
\label{eq:24}
\end{equation}
capturing the self-similar structure of the critical fluctuations in the considered system. Note that since the particle density is proportional to $\sigma^2$ (see Eq.~(\ref{eq:16})) in the numerator of Eq.~(\ref{eq:24}) occurs $\delta -1$ instead of $\delta$. In terms of the complex zeros of the partition function in Eq.~(\ref{eq:19}) the mean number of scalar particles within volume $V$ at $\zeta_c=(1,0)$ is given by Eq.~(\ref{eq:11}). Thus, according to the conjecture stated previously, there is a subset of partition function zeros $\rho_j(M)$ following the condensation law:
\begin{equation}
\rho_j(M)=1 + \gamma_j M^{-\frac{\delta - 1}{\delta +1}}
\label{eq:25}
\end{equation}
with $\gamma_j$ independent of $M$. It is straightforward to calculate also higher moments of the scalar particle multiplicity distribution for $\zeta=1$ obtaining:
\begin{equation}
\frac{\langle (\delta N)^k \rangle}{\langle N \rangle} \sim M^{d_F (k-1)/3}~~~~;~~~~\delta N = N - \langle N \rangle,~~~k=2,~3,...
\label{eq:26}
\end{equation}
In particular for $k=2$ and on the basis of the fluctuation-dissipation relation: $\chi=\frac{1}{V}\langle (\delta N)^2 \rangle$, we find the volume dependence of susceptibility at $T=T_c$: 
$$\chi_c \sim V^{\gamma/d \nu},$$
a characteristic finite-size scaling behaviour, valid in second order phase transitions ($\gamma/d \nu=2 q -1$).

For $\zeta \neq \zeta_c$ the integral in Eq.~(\ref{eq:20}) can be estimated in the saddle point approximation (since we are interested for the behaviour in the limit $M \to \infty$). Introducing the following notations:
\begin{eqnarray}
\zeta = \vert \zeta \vert e^{\mathbf{i} \theta}~~~&;&~~~\ln \zeta =\vert \ln \zeta \vert e^{\mathbf{i} \phi} \nonumber \\
\vert \ln \zeta \vert = \sqrt{(\ln \vert \zeta \vert)^2 + \theta^2}~~~&;&~~~
\phi =\rm{atan}(\frac{\theta}{\ln \vert \zeta \vert})
\label{eq:27}
\end{eqnarray}
the saddle-points in the complex $x$-plane are given by:
\begin{eqnarray}
x_k&=&M \left(\frac{2 \vert \ln \zeta \vert}{\alpha_c (\delta +1)}\right)^{\frac{2}{\delta - 1}} \exp[i \arg(x_k)] \nonumber \\\arg(x_k)&=&\frac{2 \phi}{\delta -1} + \frac{4 \pi (k-1)}{\delta -1}~;~k=1,..,k_{max}
\label{eq:28}
\end{eqnarray}
where $k_{max}=\left[\frac{4}{\delta -1}\right] + 1$ and $[..]$ means the integer part.

In terms of $x_k$ the partition function in Eq.~(\ref{eq:20}) is written:
\begin{eqnarray}
\mathcal{Z}_{sp}&=&\sum_{k=1}^{k_{max}} \sqrt{\frac{2 M \pi}{\vert F''(x_k) \vert}} \exp[M F(x_k)] \cdot \nonumber \\
&&\exp[ \mathbf{i} (\pi -\arg(F''(x_k)))/2] 
\label{eq:29}
\end{eqnarray}
with
\begin{equation}
F(x)=x \ln \zeta - a_c x^{(\delta + 1)/2}
\label{eq:30}
\end{equation}
and prime denoting derivative with respect to $x$.

To illustrate how the condensation mechanism described above works in practice, we have numerically calculated the complex zeros of the partition function (\ref{eq:19}) for $\delta = 5$ and $\alpha_c=1$ with increasing $M$. The pattern followed by these zeros is clearly displayed in Fig.~1 using $M=80$. We observe the emergence of two domains with zeros, one lying on a part of a circle and the other lying on a part of a cardioid. The latter can be identified with the critical subset of the partition function zeros while the former constitutes the complementary part. The red line indicates an analytical estimation of the critical subset employing the saddle point approximation to handle the integral in Eq.~(\ref{eq:20}), while the green line is the saddle point approximation result for the complementary subset. As $M$ increases the red arc, containing the critical subset, shrinks while the green line approaches the circumference of the unit circle (shown with black line in Fig.~1). 

\begin{figure}[tbp]
\centering
\includegraphics[width=0.45\textwidth]{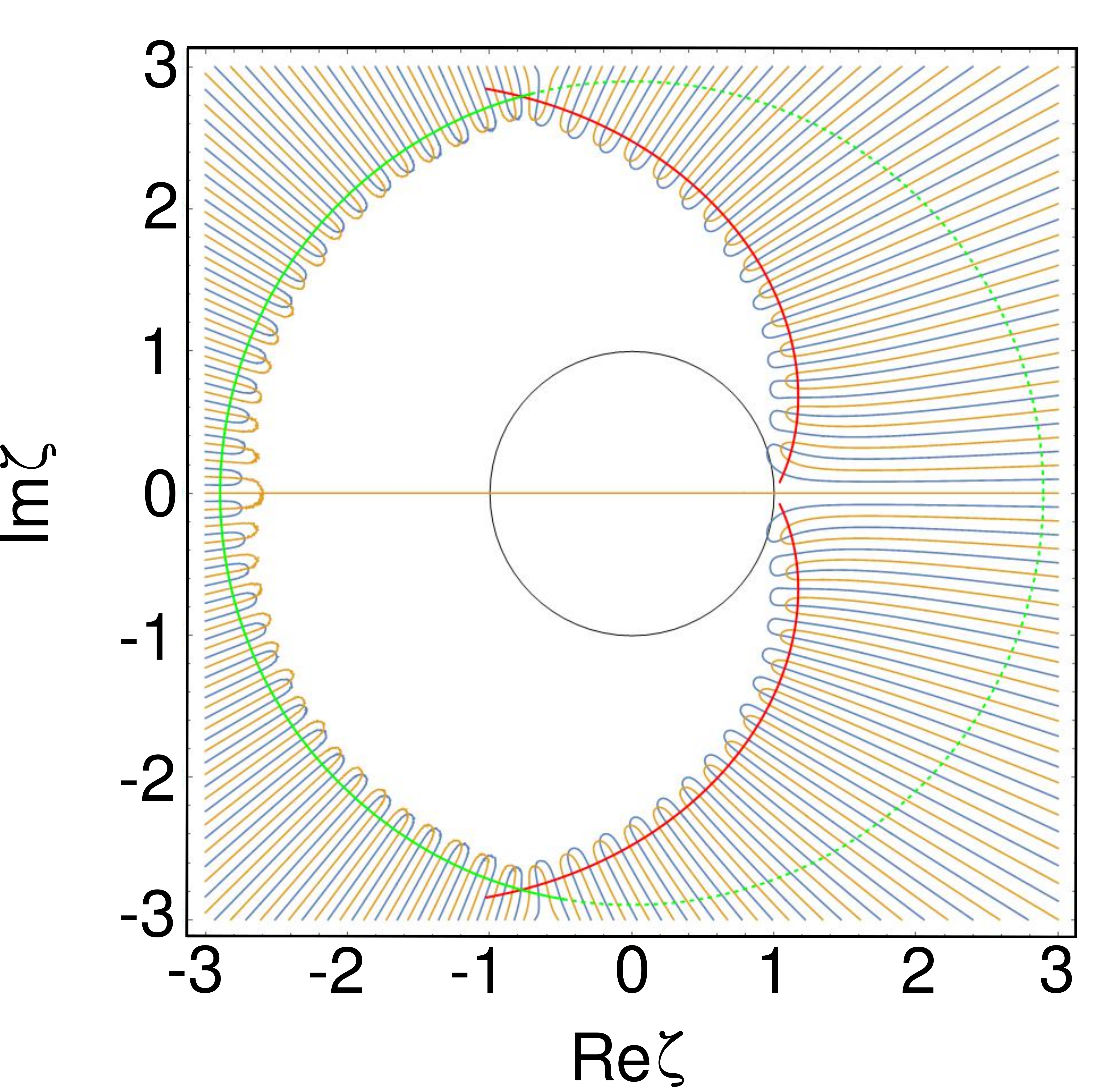}
\caption{The zeros of the partition function $\mathcal{Z}_M$ given by Eq.~(\ref{eq:19}) in the complex fugacity plane for $M=80$, 
$\alpha_c=1$ and $\delta=5$. The red and green lines are the analytical results of the saddle point approximation to the integral form of $\mathcal{Z}_M$ given by Eq.~(\ref{eq:20}). The unit circle is presented with the black line. The exact zeros lie on the intersection of the lines obeying $\Re{\mathcal{Z}_M}=0$ (blue) with the lines obeying $\Im{\mathcal{Z}_M}=0$ (orange).}
\label{fig:roots}
\end{figure}   

\begin{figure}[tbp]
\centering
\includegraphics[width=0.65\textwidth]{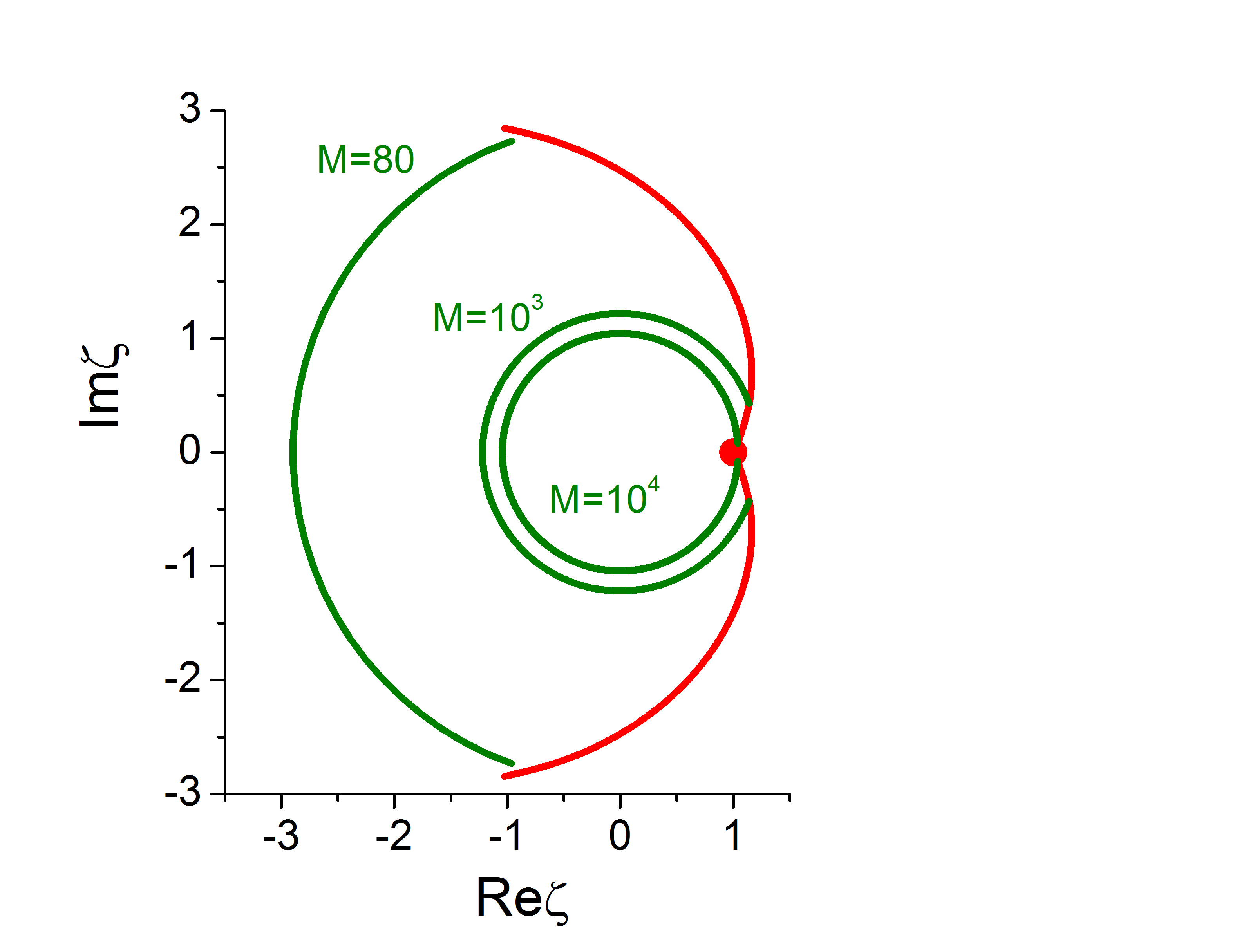}
\caption{The zeros (red and green lines) of the partition function saddle-point approximation $\mathcal{Z}_{sp}$ given in Eq.~(\ref{eq:31}) in the complex fugacity plane. We use $M=80,~1000,~10000$, $\alpha_c=1$ and $\delta=5$. The red arcs define the critical set of zeros which shrinks to the single point $\zeta_c=(1,0)$ for $M \to \infty$.}
\label{fig:saddle}
\end{figure}

We observe that for large $M$ the saddle point approximation captures sufficiently the behaviour of the roots in the complex fugacity plane. In fact for $\delta=5$ we find two saddle-points ($k_{max}=2$) obtained by setting $k=1$ and $k=2$ in Eq.~(\ref{eq:28}). Then the saddle-point approximation for the partition function in Eq.~(\ref{eq:29}) yields:
\begin{eqnarray}
\mathcal{Z}_{sp}&=&\sqrt{\frac{M \pi \exp[-\mathbf{i}\frac{\phi}{2}]}{\sqrt{3 \alpha_c \vert \ln \zeta \vert}}} \left\lbrace  \exp[\Lambda (\cos(\frac{3 \phi}{2}) + \mathbf{i} \sin(\frac{3 \phi}{2})] \right.\nonumber \\
&-& \left. \mathbf{i} \exp[-\Lambda (\cos(\frac{3 \phi}{2}) + \mathbf{i} \sin(\frac{3 \phi}{2})]\right\rbrace 
\label{eq:31}
\end{eqnarray}
with:
\begin{equation}
\Lambda=2 M \alpha_c \left(\frac{\vert \ln \zeta \vert}{3 \alpha_c}\right)^{3/2}
\label{eq:32}
\end{equation}
The zeros $\rho=\vert \zeta_{\rho} \vert \exp(\mathbf{i} \theta_{\rho})$ of the partition function (\ref{eq:29}) are given by:
\begin{eqnarray}
\cos(\frac{3 \phi_{\rho}}{2}) &=& 0~;~\phi_{\rho}=\frac{(2 \ell + 1)\pi}{3}~~,~~\ell=0,~1,~2 \nonumber \\
\tan \Lambda_{\rho} &=& 1~;~\vert \ln \zeta_{\rho} \vert = 3 \alpha_c \vert \cos \phi_{\rho} \vert \left(\frac{\pi (1 + 8 m)}{8 M \alpha_c}\right)^{2/3}
\nonumber \\
\theta_{\rho} &=& \sin \phi_{\rho} \vert \ln  \zeta_{\rho} \vert~~~(\phi_{\rho} \neq \pi)
\label{eq:33}
\end{eqnarray}
with $m \in \mathbb{N}_0$. We obtain three branches for the parametric curve defining the location of these zeros in the complex $\zeta$-plane (for $M \gg 1$). Each branch corresponds to a different value of $\ell$. In Fig.~\ref{fig:saddle} we plot this curve for $M=80,~1000,~10000$ using $\alpha_c=1$. The red arcs correspond to $\ell=0$ ($\Im \zeta > 0$) and $\ell=2$ ($\Im \zeta < 0$) while the green line corresponds to $\ell=1$. In the latter case the angle $\theta_{\rho}$ becomes independent of $\phi_{\rho}$ and $\vert \zeta_{\rho} \vert$.  

Along the red arcs, both the radial distance as well as the polar angle of the zeros scale with 
$M^{-2/3}$, verifying the proposed condensation law $\rho = 1 + \gamma M^{-2/3}$ obtained setting $\delta=5$ in Eq.~(\ref{eq:25}). For the zeros located on the green line the radial distance scales in the same way with $M$ ($\sim M^{-2/3}$) while $\theta_{\rho}=\frac{2 \pi n}{M}$ with $n \in \mathbb{N}_0$. The critical subset of the zeros corresponds to the red arcs. In Fig.~\ref{fig:saddle} the shrinking of the corresponding red arcs with increasing $M$ is clearly displayed. An additional interesting property is that the number of partition function zeros in the critical subset remains constant for sufficiently large $M$ and at the same time the length of the red arc shrinks with the law $\ell(M) \sim M^{-q}$. This implies that the linear density of the partition function zeros on the red line increases as: $P_{\rho}(M) \sim M^q$, which in turn leads to a singular measure at $\zeta=(1,0)$ for $M \to \infty$. In contrary, the number of zeros on the green line increases with $M$ following a uniform distribution on the circle's circumference.

In Fig.~3 we show the first $6$ zeros in the critical subset with the smallest distance to the critical point $\zeta_c=(1,0)$ for various values of $M$ up to $M=1000$ (green points). At the scale of this plot the zeros for $M > 1000$ are indistinguishably close to the green line and for illustrative reasons are left out.
\begin{figure}[tbp]
\centering
\includegraphics[width=0.5\textwidth]{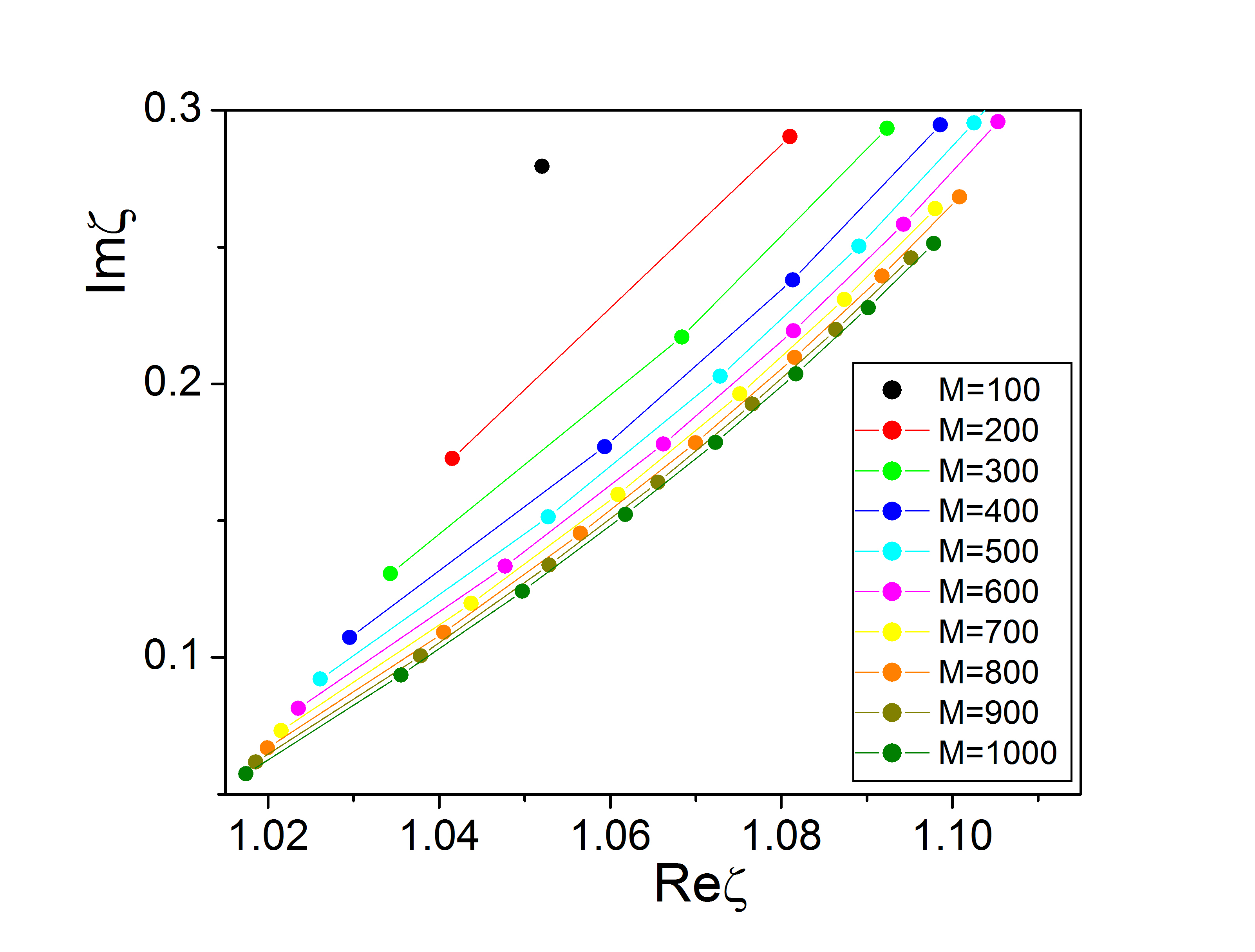}
\caption{The zeros (coloured points) of Eq.~(\ref{eq:12}) in the square $\Re \zeta \in [1,1.12]$, $\Im \zeta \in [0,0.3]$ for $\alpha_c=1$, $\delta=5$ and $M=100,~200,~..,~1000$. The solid lines are drawn to guide the eye.}
\label{fig:zeta_plane}
\end{figure}
Finally, in Fig.~4 we present in log-log plot the real and imaginary part of $\rho_j(M)-1$ as a function of $M$ for the zeros displayed in Fig.~3. A linear fit is shown for the zero closest to $\zeta_c$. The corresponding slopes are found to be $0.65$ (real part) and $0.68$ (imaginary part) very close to the theoretically predicted $q=2/3$.  

\begin{figure}[tbp]
\centering
\includegraphics[width=0.5\textwidth]{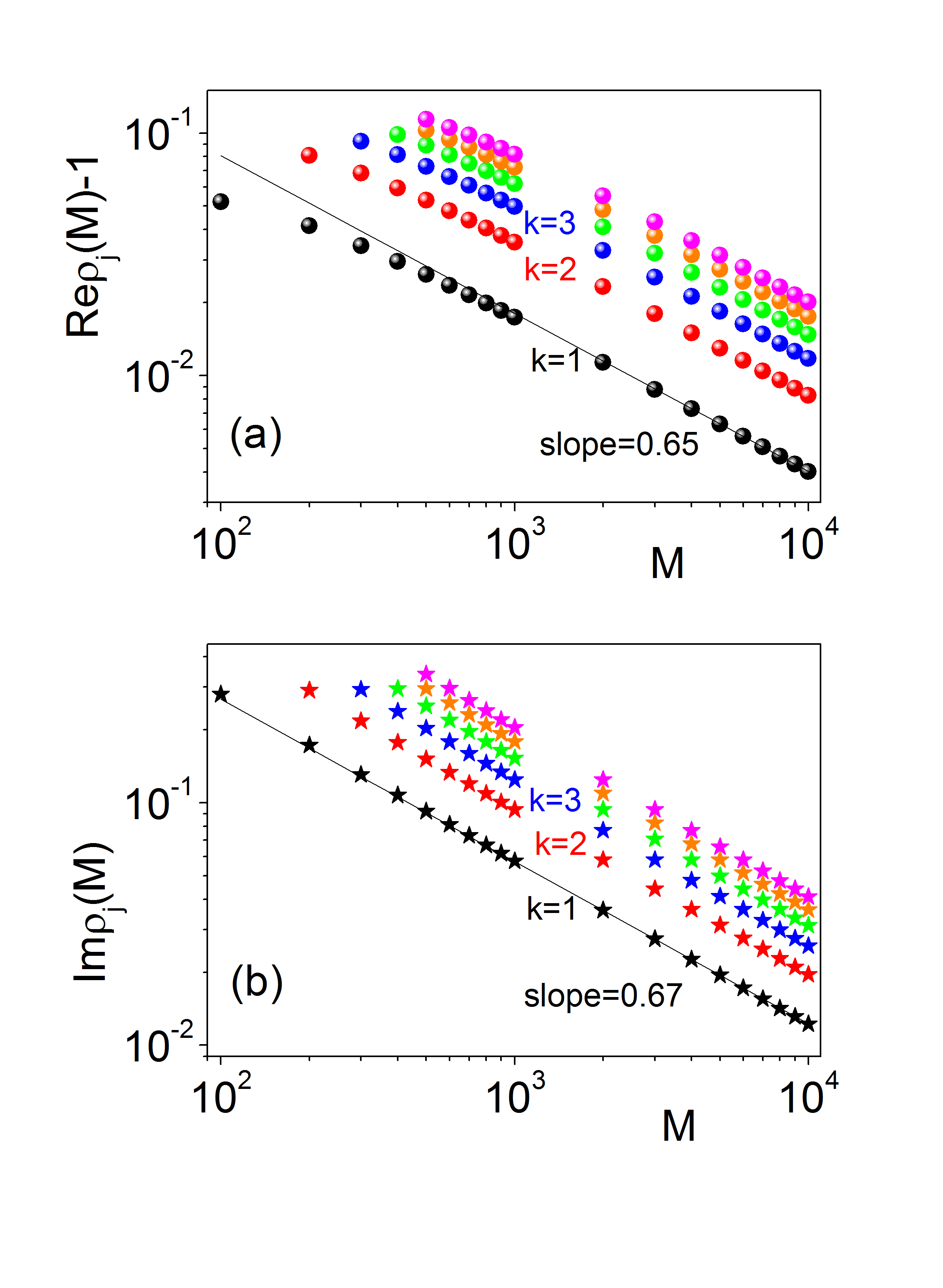}
\caption{(a) The real and (b) the imaginary part of $\rho_k(M) - 1$ as a function of $M$ for $M=100,~200,~..,~1000,~2000,~..,~10000$ and $k=1,..,6$. In both plots the black line indicates the result of a linear fit in the large $M$ ($M >4000$) range for the zero with $k=1$. As in Fig.~3 we use $\alpha_c=1$ and $\delta=5$.}
\label{fig:rootsk}
\end{figure}   


In conclusion, we have argued that the critical properties of a system, described by a partition function $\mathcal{Z}(\zeta,V,T_c)$ at $T=T_c$, are revealed by a self-similar subset of the Lee-Yang zeros $\rho_j(M)$ in the complex fugacity plane. These zeros accumulate on the region $\Re \zeta \gtrsim 1$ with increasing system size $V \sim M$ condensing to the point $\zeta_c=(1,0)$ in the limit $V \to \infty$. They capture, for any large but finite size, the most divergent part of the critical fluctuations, leading, presumably, to a transparent formulation of finite-size scaling. In particular the exponent of the particle multiplicity scaling law in Eq.~(\ref{eq:23}) is directly related to the critical fractal mass dimension $d_F$, which in turn is linked to the finite-size scaling exponent of the susceptibility at $T=T_c$. Within this framework, one may 
proceed to a systematic treatment of multiplicity measurements in experiments with relativistic nuclei in order to reconstruct, along the lines presented in \cite{Nakamura2016}, the partition function of strongly interacting matter near the QCD critical point. In this approach, the behaviour of the phenomenological Lee-Yang zeros may decide about the location of the critical point in the phase diagram, a fundamental issue of strong interactions, but also it may verify the development of a first order  or a crossover transition, when we depart from the critical point.

\end{document}